\documentstyle[aps,prb,twocolumn,epsf,floats]{revtex}
\begin{document}
%% The following two lines should be there when using 'twocolumn'.

\twocolumn[\hsize\textwidth\columnwidth\hsize\csname
@twocolumnfalse\endcsname

\title{Magnetoresistance and electronic structure of asymmetric
$\rm \bf GaAs/Al_{0.3}Ga_{0.7}As$ double quantum wells in
the in-plane/tilted magnetic field}

\draft
\author{O.~N.~Makarovskii$^{1,2}$, L.~Smr\v{c}ka$^1$,
P.~ Va\v{s}ek$^1$, T.~Jungwirth$^{1,3}$,
M.~Cukr$^1$, L.~Jansen$^4$}

\address{$^{1}$Institute of Physics ASCR, Cukrovarnick\'a 10, 162 53 Praha
6, Czech Republic\\
$^2$Institute of Radiophysics and
Electronics, NAS Ukr.\ Kharkov, Ukraine\\
$^3$ Department of Physics, Indiana University, Bloomington,
Indiana 47405\\
$^4$Grenoble High Magnetic Field Laboratory, BP 166, 38042
Grenoble Cedex 09, France}

\date{\today}
\maketitle

\begin{abstract} 
Bilayer two-dimensional electron systems formed by a thin barrier in
the GaAs buffer of a standard heterostructure were investigated by
magnetotransport measurements. In magnetic fields oriented parallel to
the electron layers, the magnetoresistance exhibits an oscillation
associated with the depopulation of the higher occupied subband and
the field-induced transition into a decoupled bilayer. Shubnikov-de
Haas oscillations in slightly tilted magnetic fields allow to
reconstruct the evolution of the electron concentration in the
individual subbands as a function of the in-plane magnetic field. The
characteristics of the system derived experimentally are in {\em
quantitative} agreement with numerical self-consistent-field
calculations of the electronic structure.
\end{abstract}

\pacs{PACS numbers: 73.20.Dx, 73.40.-c, 73.50.-h}
%\date{\today}
%% The following line should be there when using 'twocolumn'.

\vskip2pc] 
\section{Introduction} 
\label{intro} 
In an idealized, infinitely narrow two-dimensional (2D) system the
in-plane component of the magnetic field couples only to the
electronic spin degree of freedom.  In real samples, however, the
finite size of a 2D system in the growth ($\hat{z}$) direction often
leads to strong orbital effects of the in-plane field.  For example,
2D-2D tunneling studies\cite{eis} of a system consisting of two nearby
narrow quantum wells, which is the simplest structure with non-trivial
growth direction degree of freedom, have revealed dramatic effects of
the in-plane field on the Fermi surface topology.  In this
weakly-coupled double quantum well sample the in-plane field,
$B_{\parallel}$, in effect displaces the origin of the two layer Fermi
circles by $|e|B_{\parallel}d/\hbar$ and allows a sweep of one past
the other at a critical field $B_c=2\hbar k_F/|e|d$, where $d$ is the
layer separation and $k_F$ is the individual layer Fermi wavevector.

The crossing of Fermi surfaces is replaced with more complicated
patterns in samples with strongly coupled quantum wells in which the
tunneling rate dominates the electronic scattering rate within a well.
In a simple tight-binding model\cite{hu} a partial energy gap, equal
to the bonding-antibonding gap at zero field, opens at the wavevector
corresponding to the crossing point of the displaced dispersion curves
for the uncoupled layers. While the upper subband above the energy gap
maintains a nearly parabolic shape, a saddle point develops in the
lower energy subband.  With increasing $B_{\parallel}$ the bottom of
the upper subband moves above the Fermi energy at a critical field
$B_{c,1}$, resulting in a sudden drop in the density of states. At the
second critical field, $B_{c,2}$, the saddle point of the lower
subband approaches the Fermi energy and the density of states
diverges. The magnetoresistance oscillation observed on coupled
double\cite{si,ku} and triple\cite{lay} quantum wells represents a
striking manifestation of the two distinct van Hove singularities in
the $B_{\parallel}$-dependent density of states.  For these samples
the tight-binding model has provided an accurate quantitative estimate
for the critical in-plane fields.  Bilayer systems realized in wide
single quantum wells also display\cite{ju} the magnetoresistance
oscillation, however, the orbital effects of the in-plane magnetic
field are more complex here and cannot be captured by the simple
tight-binding approximation. Instead, the numerical
self-consistent-field technique has proven successful\cite{ju} for
wide quantum wells.

The importance of orbital effects of the in-plane magnetic field has
recently been emphasized\cite{dassarma} also in the context of
metal-insulator transition studies in Si-MOSFET\cite{mit} and
GaAs/Al$_{0.3}$Ga$_{0.7}$As single heterojuncions.\cite{yoon,papa} The
observed dramatic response to $B_{\parallel}$ in both metallic and
insulating phases has been attributed\cite{dassarma,papa} to the spin
coupling to in-plane field as well as to the
distortion\cite{zaremba,ussingle} of carrier Fermi surfaces. A
detailed quantitative understanding of these phenomena requires to
incorporate the orbital effects of $B_{\parallel}$ using
non-perturbative approaches.

In this paper we present a study of magnetotransport properties of
GaAs/Al$_{0.3}$Ga$_{0.7}$As heterojunctions with an additional thin
Al$_{0.3}$Ga$_{0.7}$As barrier introduced into the GaAs
buffer.\cite{sm} The structures were specially designed to combine
properties of single-junction and bilayer systems. A comparison
between theory and experiment in our samples constitutes an excellent
test for the reliability of the numerical self-consistent-field
technique applied to systems with complex growth direction
geometries. In Section II we present measured resistance oscillation
with in-plane field and Shubnikov-de Haas (SdH) oscillations recorded
at magnetic fields slightly tilted from the 2D layer plane.
Theoretical calculations of the $B_{\parallel}$-dependent electronic
structure are used, in Section III, for quantitative interpretation of
the experimental data.  Our conclusions are summarized in Section IV.

%*************************************************************************
\section{Experimental results}
\begin{figure}[htb]
\leavevmode
\begin{center}
\epsfxsize = 75 mm \epsfbox{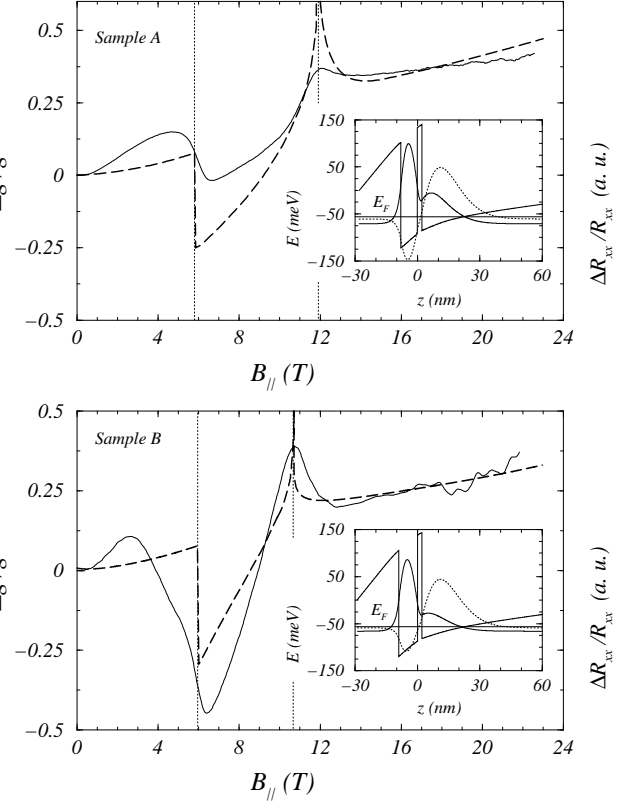}
\end{center}
\caption{ The magnetoresistance trace (solid line) recorded for
samples A and B at in-plane magnetic fields. For both samples the
magnetoresistance curve is multiplied by a constant to fit the
calculated density of states (dashed line) at high $B_{\parallel}$.
The insets show the band-profile and the wavefunctions of the bonding
and antibonding subbands at $B = 0$. The wavefunctions are shifted in
the vertical direction so that their asymptotic values match the
corresponding quantum levels on the energy axis. In both samples, the
levels lie below the Fermi energy, $E_F$, indicating two occupied
subbands at zero field.}
\label{fig1}
\end{figure}

We studied modulation-doped GaAs/Al$_{0.3}$Ga$_{0.7}$As
heterojunctions grown by molecular-beam epitaxy on (100) oriented
semi-insulating GaAs substrates and patterned in the standard Hall-bar
samples. A thin barrier formed by eight monolayers of
Al$_{0.3}$Ga$_{0.7}$As was grown inside the GaAs buffer producing two
coupled quantum wells, one of a distorted rectangular shape and the
other one with a nearly triangular geometry, as shown in  insets of
Fig.\ \ref{fig1}. We chose the thickness and position of the barrier to
achieve high population of the second (antibonding) subband. The
particular structure design was based on zero-field numerical
self-consistent simulations that predict minimum bonding-antibonding
gap for barrier positioned near the node of the first excited state
wavefunction (peak of the lowest energy state wavefunction) of the
original single-junction. Note that unlike the conventional bilayers
realized in double quantum wells or wide single wells our system is
intrinsicly, strongly anisotropic. At zero magnetic field the lower
(bonding) subband wavefunction has dominant weight in the rectangular
well while electrons from the higher (antibonding) subband are more
likely to occupy the triangular quantum well.  Parameters of the two
studied samples are summarized in Table~\ref{samples}.  The partial
occupations of the bonding and antibonding subbands at
$B_{\parallel}=0$ were obtained by Fourier analysis of the low-field
SdH oscillations. The Hall measurement at low perpendicular magnetic
fields provided an independent check for the total carrier density and
was used to determine sample mobilities.  The insets in
Fig.\ \ref{fig1} show the confining potential profiles and wavefunctions
of the occupied subbands for two samples.

\begin{table}[h]
\leavevmode
\begin{center}
\begin{tabular}{ccccccc}
Sample & $w$ & $\mu$ & $N_b$ & $N_a$ & $N_{Hall}$ & $\overline{N}$ \\
& ($\AA$) & (10$^5$ cm$^2$/Vs) & \multicolumn{4}{c}{(10$^{11}$
cm$^{-2}$)} \\ \hline A & 80 & 2.00 & 2.46 & 0.89 & 3.38 & 3.4\\ B &
90 & 2.34 & 2.77 & 0.76 & 3.42 & 3.5
\end{tabular}
\end{center}
\caption{\protect Sample parameters: $w$ is the thickness of the
rectangular well, $N_b$ and $N_a$ are the bonding and antibonding
subband densities obtained from SdH measurement, $N_{Hall}$ is the
total 2D electron density derived from low-field Hall data, and
$\overline{N}$ is the average between the SdH and Hall values for the
total density used in the numerical simulations.}
\label{samples}
\end{table}

\begin{figure}[htb] 
\leavevmode
\begin{center} 
\epsfxsize = 75 mm \epsfbox{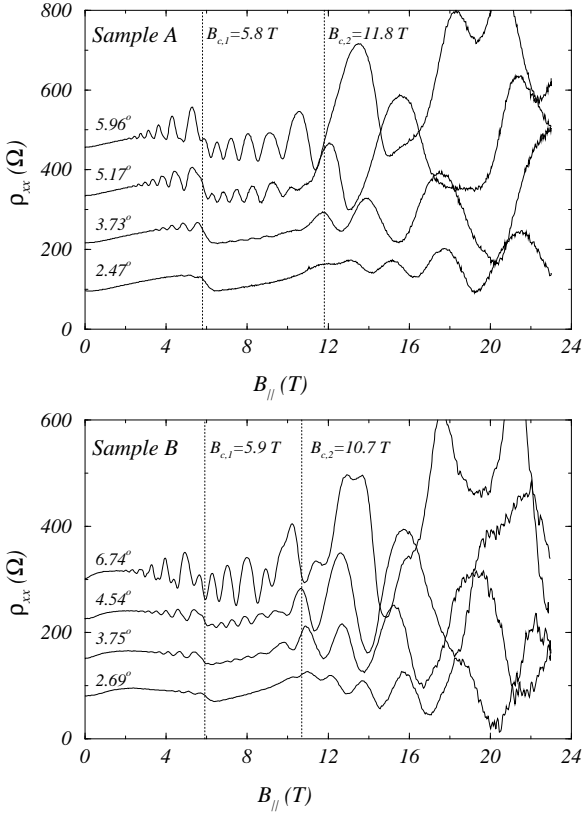}
\end{center} 
\caption{Typical magnetoresistance traces measured for small angles
between the sample plane and the magnetic field direction.  The curves
for higher angles are shifted upward by 75 $\Omega$.}
\label{fig2} 
\end{figure}

\begin{figure}[htb]
\leavevmode
\begin{center}
\epsfxsize = 75 mm \epsfbox{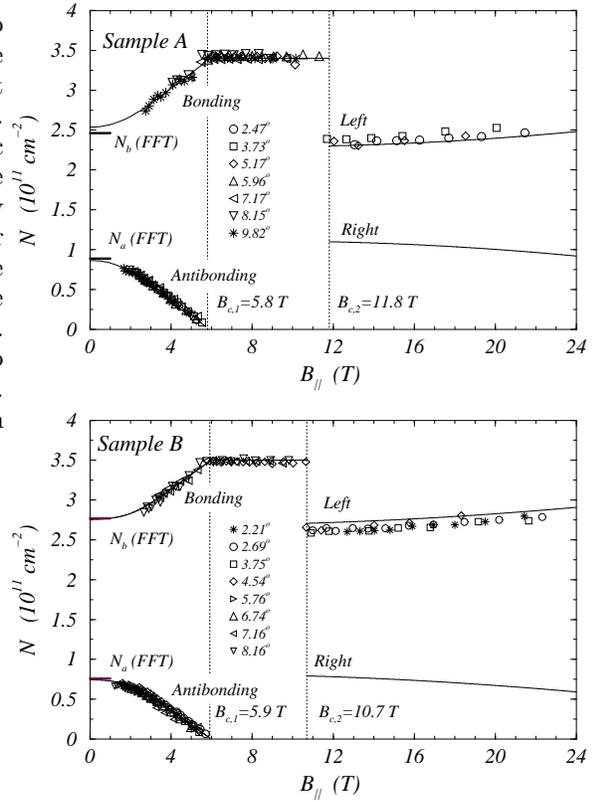}
\end{center}
\caption{The 2D electronic concentrations in occupied subbands.  Solid
lines represent theoretical results, marked points denote data derived
from magnetoresistance measurements.  Due to strongly field-dependent
cyclotron effective masses, measurements at different groups of angles
were appropriate for different subbands and different in-plane field
regions. $N_a$ and $N_b$ denote experimental results obtained from the
Fourier transformation of Shubnikov-de Hass oscillations measured at
low perpendicular magnetic fields ($B <$ 1).}
\label{fig3}
\end{figure}

Magnetotransport data were collected at the temperature 0.45 K and in
magnetic fields ranging from 0 to 23~T, using both dc and
low-frequency (13 Hz) ac techniques. First, we measured the resistance
$R_{xx}$ for a magnetic field precisely parallel with the 2D layer
plane. The recorded magnetoresitance, $\Delta R_{xx}/R_{xx} =
(R_{xx}(B_{\parallel}) - R_{xx}(0))/R_{xx}(0)$, is plotted in
Fig.\ \ref{fig1}. Consistent with the bilayer nature of the studied
samples we observe an oscillation on the magnetoresistance trace with
the lower critical field $B_{c,1}$ and the upper critical field
$B_{c,2}$. The values of critical fields are $B_{c,1}=5.8$~T,
$B_{c,2}=11.7 $~T for sample A and $B_{c,1}=5.9$~T, $B_{c,2}=10.7 $~T
for sample B, respectively.  The in-plane magnetic field dependence of
subband occupations, $N_a(B_{\parallel})$ and $N_b(B_{\parallel})$,
were obtained from SdH oscillation data measured at different
field-tilt angles. Typical SdH traces are shown in Fig.\ \ref{fig2}. In
general, the oscillations are not periodic in $1/B_{\perp}$
($B_{\perp}$ is the perpendicular component of the magnetic field)
since both $B_{\perp}$ and $B_{\parallel}$ vary when the field is
swept at a fixed tilt angle.
Therefore, the standard Fourier technique does not
apply. Instead, we use the measured distance between valleys
surrounding individual peaks to identify subband densities.\cite{fa}
For $B_{\parallel}<B_{c,1}$ one type of oscillations, corresponding
to the lower density (antibonding) subband, can be detected at small
tilt angles while two subbands are clearly visible at higher
angles. With increasing $B_{\parallel}$ the distribution of electrons
between the two subbands changes, as shown in Fig.\ ~\ref{fig3}.
Eventually, the antibonding subband is depopulated at
$B_{\parallel}=B_{c,1}$.  At higher in-plane fields two regimes can be
distinguished.  For $B_{c,1}~<B_{\parallel}<B_{c,2}$ the SdH
oscillations are periodic indicating a single occupied
subband. Consistently, the corresponding density is equal to the total
2D density measured at zero in-plane magnetic field. The character of
the oscillations changes abruptly at $B_{\parallel}=B_{c,2}$; the low
in-plane field non-periodicity is recovered and the obtained carrier
density is significantly reduced (see Fig.\ \ref{fig3}). These results
suggest that the Fermi sea splits into two disconnected parts of
densities that vary with $B_{\parallel}$. Only one Fermi surface,
however, can be detected from our SdH data.

%************************************************************************
\section{Theoretical analysis}
The single-particle Hamiltonian for an electron confined in a
GaAs/Al$_{0.3}$Ga$_{0.7}$As heterostructure and subjected to in-plane
magnetic field can be written as
\begin{equation}
\label{ham}
H = \frac{1}{2m^*}({\bf p} + |e|{\bf A})^2 + V_{conf}(z),
\end{equation}
where $m^*$ is the effective mass in the GaAs conduction band and the
vector potential for $B_{\parallel}$ applied along the $\hat
y$-direction takes a form ${\bf A} = (B_{\parallel}z,0,0)$. The
confining potential $V_{conf}(z)$ is constructed using the nominal
growth parameters and includes Hartree and exchange-correlation
potentials generated by the free carriers in the quantum wells. The
Hartree potential is derived from the $z$-dependent density of
electrons, $\rho(z)$, by numerical solution of the Poisson
equation. The exchange-correlation term is calculated within the
local-density approximation.\cite{vosko} In each loop of the
self-consistent procedure we solve numerically the Schr\"{o}dinger
equation with the Hamiltonian (\ref{ham}) to get $\rho(z)$. Then, a
new $V_{conf}(z)$ is constructed, which enters the next loop of the
procedure until the self-consistency condition has been achieved. The
resulting energy spectra have the form
\begin{equation}
E_i({\bf k}) = E_{i,x}(k_x) + \frac{\hbar^2}{2m}k_y^2,
\end{equation}
where $E_{i,x}(k_x)$ depends on the sample geometry and on the
magnitude of the in-plane magnetic field. An index $i=a,b$
distinguishes the antibonding and bonding subband states.

2D subband concentrations $N_i$ are proportional to the area enclosed by
corresponding Fermi contours.  The partial densities of states (DOS) $g_i$
and the cyclotron effective masses $m_{c,i}$ are related to the shape of
the Fermi contours by the following expression: 
\begin{equation}
\label{mc}
      g_i = \frac{m_{c,i}}{\pi \hbar^2} = \frac{1}{2\pi^2}
          \oint \frac{dk}{|\nabla_kE_i|}.
\end{equation}
The total density of states $g = g_a + g_b$.  The distortion of Fermi
lines can be probed experimentally by adding a weak perpendicular
component of the magnetic field. For $B_{\perp}\ll B_{\parallel}$, the
quantization of the in-plane component of the electron motion can be
described in terms of quasiclassical Landau levels with the cyclotron
effective mass given by Eq.\ (\ref{mc}).  The degeneracy of the
spin-unresolved levels is $2|e|B_{\perp}/h$. 

\begin{figure}[htb]
\leavevmode
\begin{center}
\epsfxsize = 75 mm
\epsfbox{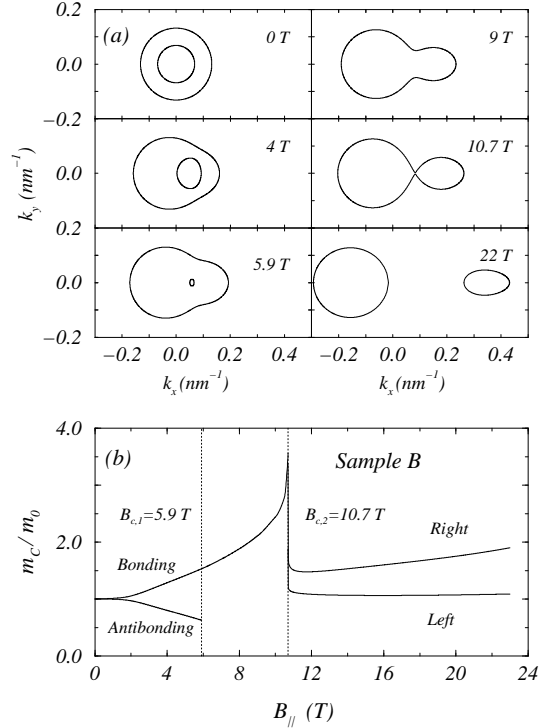}
\end{center}
\vspace{-10mm}
\caption{(a) Theoretical evolution of  Fermi
contours in the in-plane magnetic field. (b) Self-consistently calculated 
cyclotron effective masses. }
\label{fig4}
\end{figure}

The evolution of Fermi surfaces in $B_{\parallel}$ is illustrated in
Fig.\ \ref{fig4} for sample B. Fig.\ \ref{fig4}(a) presents theoretical
Fermi contours calculated for several selected values of
$B_{\parallel}$, corresponding cyclotron effective masses are shown in
Fig.\ \ref{fig4}(b).  The deviations from the zero-field circles
reflect the field-induced changes of $E_{i,x}(k_x)$ described for the
case of the simple tight-binding model in Section~\ref{intro}. The
antibonding $E_{a,x}(k_x)$ is an asymmetric function of $k_x$,
narrower than the free-electron parabola, with the minimum (bottom)
moving to the Fermi energy which crosses at $B_{\parallel}=B_{c,1}$.
The corresponding Fermi contour acquires the shape of a ``lens'',
i.e. the oval with the longer axis oriented in the
$\hat{k}_y$-direction. Due to this type of deformation the antibonding
cyclotron effective mass $m_{c,a}$ is a decreasing function of
$B_{\parallel}$.

The field-induced local maximum developing in the $k_x$-dependence of
$E_{b,x}(k_x)$ causes elongation of the bonding Fermi contour in the
$\hat{k}_x$-direction. For higher fields, the Fermi line evolves into
an asymmetric ``peanut'' shape.  A neck connecting left and right
parts of the ``peanut'' breaks as the saddle point reaches the Fermi
energy at $B_{\parallel}=B_{c,2}$. For $B_{\parallel}<B_{c,2}$ the
bonding effective mass $m_{c,b}$ grows with $B_{\parallel}$ and
diverges at the saddle point. Above $B_{c,2}$, the Fermi contour
splits into two approximately elliptic lines. Very similar results are
obtained for sample B.

The position of an electron described by the state $|i,k_x,k_y\rangle$
in one or the other well is given by $\langle z\rangle_{i,x} =\langle
i,k_x,k_y|z|i,k_x,k_y\rangle$. The coordinate $\langle z\rangle_{i,x}$
is related to the $k_x$-component of the wave vector by
\begin{equation}
     \langle z\rangle_{i,x}  =  \frac{\hbar k_x}{m\omega}
     -\frac{\langle v\rangle_{i,x}}{\omega},
\end{equation}
\begin{equation}
       \langle v\rangle_{i,x}  =  \frac{1}{\hbar}\frac{\partial 
      E_{i,x}(k_x)}{\partial k_x},
\end{equation}
where $\omega = |e|B_{\parallel}/m^*$ and $\langle v\rangle_{i,x}$ is
the $x$-component of the in-plane group velocity. These relations
imply that the larger, almost circular Fermi contour describes the
electron layer in the left rectangular well while the smaller
elongated oval corresponds to states in the right triangular well.
The cyclotron effective masses in two wells are different: above
$B_{c,2}$ the mass for the left well drops quickly close to the
zero-field value $m^*$, the mass for the triangular well also drops,
but to substantially larger value.

\begin{figure}[htb]
\leavevmode
\begin{center}
\epsfxsize = 75 mm
\epsfbox{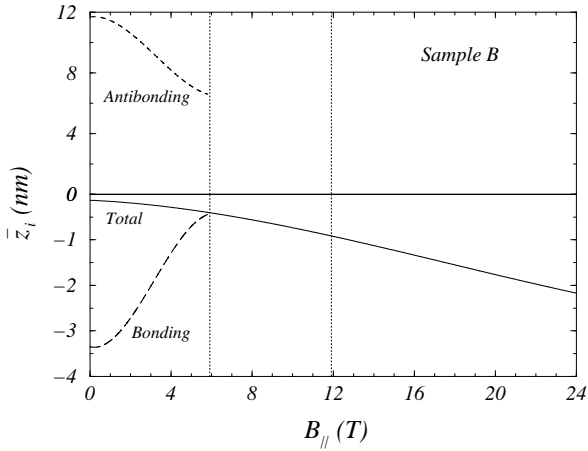}
\end{center}
\vspace{-5cm}
\caption{Theoretical centroids $\bar z_i$, $i=a,b$, of densities
$\rho_a(z)$, $\rho_b(z)$ and their sum $\rho(z)= \rho_a(z)+\rho_b(z)$
as functions of in-plane magnetic field. Note the different scales for
the positive  and negative parts of the  vertical axes.}

\label{fig5}
\end{figure}

Both the partial densities of electrons  $\rho_a(z)$, $\rho_b(z)$ and their sum
$\rho(z)= \rho_a(z)+\rho_b(z)$ are influenced by  in-plane magnetic
fields. In our structure with a hard-wall barrier  the densities have sharp
minima in the barrier and peaks inside the wells.  The position of
peaks is almost the same for all densities and practically field-independent, 
the magnetic field effect is represented by  changes of 
relative magnitudes of the peaks.  We will characterize the transfer of
electrons between wells by $\bar z_i(B_{\parallel})$, the
field-dependent centroids of electron densities.
Results calculated for sample B are shown in Fig.\ \ref{fig5}. The
center of mass of $\rho(z)$  is in the middle of the structure
at $B_{\parallel}=0$ and moves closer to the interface as the field
increases. This behavior is typical for single-junction
structures.\cite{ei} Below $B_{c,1}$, the centroid $\bar z_a$ of
antibonding electrons is inside the triangular well for all fields. At
zero-field, $\bar z_a$ is far from the barrier. The growing $B_{\parallel}$
transfers  antibonding electrons from the triangular to rectangular
well and, consequently, $\bar z_a$ is shifted towards the barrier. The
electrons from the bonding subband exhibit opposite behavior in this
range of fields: the magnetic field empties the antibonding subband
and corresponding  electrons, mostly from the triangular well, become a
part of the bonding subband. As result, $\bar z_b$ moves from position
close to the interface to the position near the barrier as
$B_{\parallel}$ increases. Above $B_{c,1}$, when only the bonding
subband is occupied, the transfer of  electrons from the triangular
to rectangular well continues. This is the case also above $B_{c,2}$ when
the bonding Fermi contour is splitted into the left and right parts,
as seen in Figs.\ \ref{fig3} and \ref{fig4}.

The characteristic of the samples A and B derived from the
magnetoresitance experiments are collected in Figs.~\ref{fig1} and
\ref{fig3}, together with results of the self-consistent numerical
simulation. In the case when theory cannot be directly compared with
experiments only the theoretical results for sample B are presented.

We associate the sharp minima in the experimental magnetoresistance
traces, presented in Fig.\ \ref{fig1} with the sudden decrease of the
density of states at $B_{c,1}$ that contributes to reduction of the
scattering rate at the Fermi energy. The difference between the
field-dependence of cyclotron effective masses $m_{c,a}$ and
$m_{c,b}$, calculated for sample B and shown in Fig.\ \ref{fig4}(b),
explains why the SdH oscillations of electrons in the antibonding
subband are seen at lower tilt-angles than the oscillations of
electrons from the bonding subband.

The increase of theoretical $m_{c,b}$ at in-plane fields below
$B_{c,2}$ is related to the strong distortion of the Fermi contour, as
seen in Fig~\ref{fig4}.  Only the bonding subband is occupied for
$B_{c,1}<B_{\parallel}<B_{c,2}$ and, therefore, the SdH oscillations
are periodic in $1/B_{\perp}$.  The decrease of their amplitude for
$B_{\parallel}$ approaching $B_{c,2}$ is an experimental confirmation
of calculated sharp increase of $m_{c,b}$.  The theoretical value for
the critical in-plane field precisely matches with $B_{c,2}$ obtained
from experiment. The peak in the measured in-plane field
magnetoresistance at $B_{c,2}$ can be understood as a consequence of
the divergence of the DOS at Fermi energy or, in other words, of a
zero group velocity of
electrons at the saddle point.

The experimental magnetoresistance oscillation is clearly associated
with van Hove singularities in the $B_{\parallel}$-dependent
DOS. However, apart from these strong features $R_{xx}$ does not
follow $g$ for in-plane fields less than $B_{c,2}$ (see Fig.\
\ref{fig1}).  The differences between $\Delta R_{xx}/R_{xx}$ and
$\Delta g/g $ curves stem from the in-plane field suppression of the
coupling between wells, the transfer of electrons from the triangular
to rectangular well, and the changes in the nature of electronic
scattering.

At $B_{\parallel}>B_{c,2}$, the Fermi contour splits into two
approximately elliptic lines.  The larger, almost circular Fermi
contour is for the rectangular well while the smaller elongated oval
corresponds to states in the triangular well. Since electrons are
localized in one or the other quantum well, the scattering rate
becomes nearly independent of $B_{\parallel}$, and $\Delta
R_{xx}/R_{xx} \propto \Delta g/g $ applies, as seen from Fig.\
~\ref{fig1}. Large cyclotron effective mass and low concentration of
electrons in the triangular well explain while these states are not
detectable by the SdH measurement; at most two weak oscillations can
occur in available magnetic fields. The calculated density of
electrons in the rectangular well is again in excellent quantitative
agreement with SdH data.

%************************************************************************
\section{Conclusions}
We have performed an experimental and theoretical study of asymmetric
GaAs/Al$_{0.3}$Ga$_{0.7}$As heterostructure consisting of coupled
rectangular and triangular quantum wells. The positions of a strong
minimum and maximum in the measured in-plane field dependent
resistance of the 2D electron system match precisely with calculated
van Hove singularities in the DOS. Theoretical predictions for
$B_{\parallel}$-dependent occupations of electronic subbands and
quantum wells are in quantitative agreement with SdH data recorded at
magnetic fields slightly tilted from the 2D layer plane.  We conclude
that the numerical self-consistent-field technique provides a
realistic description of orbital effects of the in-plane field on 2D
electron systems confined in semiconductor heterostructures with
general growth-direction geometry.

To achieve similar level of accuracy for valence-band states with strong
spin-orbit coupling represents a formidable challenge for future
theoretical work.  Particularly intriguing is the derivation of
local-spin-density approximation for many-body states with unequal
population of different spin-subbands.  Recent remarkable studies of the
metal-insulator transition in high mobility GaAs hole systems certainly
give a strong motivation for pursuing this kind of research.

This work was supported by the Czech-French project Barrande 99011 and
by the Grant Agency of the Czech Republic under Contract No. 202/98/0085.


\begin{references}
\bibitem{eis} J.\ P.\ Eisenstein, T.\ J.\ Gramila, L.\ N.\ Pfeiffer,
and K.\ W.\ West, Phys. Rev. B {\bf 44}, 6511 (1991).

\bibitem{hu} J.\ Hu and A.\ H.\ MacDonald, Phys.\ Rev.\ B {\bf
46}, 12554 (1992).

\bibitem{si} J.\ A.\ Simmons, S.\ K.\ Lyo, N.\ E.\ Harff, and J.\
F.\ Klem,   Phys. Rev. Lett. {\bf 73}, 2256 (1994);
S.\ K.\ Lyo, Phys.\ Rev.\ B {\bf 50}, 4965 (1994).

\bibitem{ku} A.\ Kurobe, I.\ M.\ Castleton, E.\ H.\ Linfield. M.\ P.\
Grimshaw, K.\ M.\ Brown, D.\ A.\ Ritchie, M.\ Pepper, and G.\
A.\ C.\ Jones, Phys. Rev. B, {\bf 50}, 11492 (1994).

\bibitem{lay} T.\ S.\ Lay, X.\ Ying, and M.\ Shayegan,
Phys. Rev. B {\bf 52}, R5511 (1995).

\bibitem{ju} T.\ Jungwirth, T.\ S.\ Lay, L.\ Smr\v{c}ka, and M.\
Shayegan, Phys. Rev. B, {\bf 56}, 1029 (1997).

\bibitem{dassarma} S.\ Das Sarma and E.\ H.\ Hwang,
cond-mat/9909452.

\bibitem{mit} D.\ Simonian, S.\ V.\ Kravchenko, M.\ P.\ Sarachik,
and V.\ M.\ Pudalov, Phys. Rev. Lett. {\bf 79}, 2304 (1997);
T.\ Okamoto, K.\ Hosoya, S.\ Kawaji, and
a.\ Yagi, Phys. Rev. Lett. {\bf 82}, 3875 (1999).

\bibitem{yoon} Jangsoo Yoon, C.\ C.\ Li, D.\ Shahar, D.\ C.\ Tsui,
and M.\ Shayegan, Phys. Rev. Lett., in press (2000).

\bibitem{papa} S.\ J.\ Papadakis, E.\ P.\ De Poortere, and
M.\ Shayegan, cond-mat/9911239.

\bibitem{zaremba} J.\ M.\ Heisz and E.\ Zaremba, Semicond.
Sci. Technol. {\bf 8}, 575 (1993).

\bibitem{ussingle} T.\ Jungwirth and L.\ Smr\v{c}ka,
J. Phys.: Condens. Matter {\bf 5}, L217 (1993);
L.\ Smr\v{c}ka, P.\ Va\v{s}ek, J.\ Kol\'a\v{c}ek,
T.\ Jungwirth, and M.\ Cukr, Phys. Rev. B {\bf 51}, 18011 (1995).

\bibitem{sm} L.\ Smr\v{c}ka, P.\ Va\v{s}ek, T.\ Jungwirth, O.\
N.\ Makarovskii, M.\ Cukr, and L.\ Jansen, Acta Physica Polonica
A, {\bf 92} (1997).

\bibitem{fa} G.\ R.\ Facer, B.\ E.\ Kane, R.\ G.\ Clark, L.\ N.\
Pfeiffer, K.\ W.\ West,  Phys. Rev. B, {\bf 82}, R10036 (1997).

\bibitem{vosko} S.\ H.\ Vosko, L.\ Wilk, M.\ Nusair, Can. J.
Phys. {\bf 58}, 1200 (1980).

\bibitem{ei} J.\ Hampton, J.\ P.\ Eisenstein, L.\ N.\ Pfeiffer, and
K.\ W.  West, Solid State Commun.\ {\bf 94}, 559 (1995); 
T.\ Jungwirth and  L.\ Smr\v{c}ka, Phys. Rev. B, {\bf 51}, 10 181 (1995).
\end{references}
\end{document}